\documentclass[preprint,showpacs,preprintnumbers,amsmath,amssymb,nofootinbib]{revtex4}
\usepackage[dvips]{graphicx}
\usepackage{graphicx}
\usepackage{dcolumn}
\usepackage{bm}
\usepackage{axodraw}

\def\lsim{\raise0.3ex\hbox{$\;<$\kern-0.75em\raise-1.1ex\hbox{$\sim\;$}}}
\def\gsim{\raise0.3ex\hbox{$\;>$\kern-0.75em\raise-1.1ex\hbox{$\sim\;$}}}
\newcommand{\bi}{\bibitem}
\newcommand{\be}{\begin{eqnarray}}
\newcommand{\ee}{\end{eqnarray}}
\newcommand{\bea}{\begin{array}}
\newcommand{\eea}{\end{array}}
\newcommand{\nn}{\nonumber}

\newcommand{\TeV}{\mbox{TeV}}
\newcommand{\U}{\cal{U}}
\newcommand{\BZ}{\cal{BZ}}

\renewcommand{\eqref}[1]{Eq.~(\ref{#1})}
\newcommand{\figref}[1]{Fig.~\ref{#1}}

\catcode`\@=11
\catcode`\@=12

\begin{document}
\def\descriptionlabel#1{\bf #1\hfill}
\def\description{\list{}{%
\labelwidth=\leftmargin
\advance \labelwidth by -\labelsep
\let \makelabel=\descriptionlabel}}

\title{\vspace{2cm} Muon Anomalous Magnetic Moment and Lepton Flavor Violating \\ 
Tau Decay in Unparticle Physics
\vspace{1cm}}

\author{Andi Hektor,\footnote{andi.hektor@cern.ch}~Yuji Kajiyama,\footnote{yuji.kajiyama@kbfi.ee} and Kristjan Kannike\footnote{kkannike@ut.ee}}
\affiliation{National Institute of Chemical Physics and Biophysics,
Ravala 10, Tallinn 10143, Estonia
\vspace{3cm}
}

\begin{abstract}
We study effects of unparticle physics on muon $g-2$ and 
LFV tau decay processes. 
LFV interactions between the Standard Model sector and 
unparticles can explain the difference of experimental value of muon $g-2$ 
from the Standard Model prediction. While the same couplings generate 
LFV tau decay, we found that LFV coupling can be of ${\cal O}(0.1 \ldots 1)$ 
without conflict with experimental bounds of LFV tau decay
if the scaling dimension of unparticle operator $d_{\U}\gsim 1.6$. 
\vspace{2cm}
\end{abstract}


\maketitle
\section{Introduction}

Scale or conformal invariance of a quantum field theory requires that particles included in that theory be massless. 
In the Standard Model (SM), scale invariance is broken by the mass parameter in the 
Higgs sector and by running of gauge couplings. However, this does not forbid the existence of scale invariant hidden sector. 
Recently, motivated by Banks and Zaks \cite{bz}, 
Georgi suggested \cite{georgi} that there may exist a 
scale invariant hidden sector of unparticles $\U$ coupled to the Standard Model (SM) at TeV scale. 
The theory at high energy contains both the SM fields and so-called Banks-Zaks $(\BZ)$ fields of a theory with with a non-trivial infrared fixed point, interacting via messenger fields of high mass. At the TeV scale, $\BZ$ fields are mapped to  effective scale invariant unparticle operators interacting with the SM fields. An intriguing property of unparticles is their non-integral scaling dimension $d_{\U}$. They behave like $d_{\U}$ number of massless invisible particles. 

Since the principle to constrain the interactions of unparticles 
with SM sector is still unknown, there are many possibilities 
of interactions that preserve Lorentz structure: unparticles that are 
SM gauge singlets \cite{chen},  have baryon number \cite{he} or gauge quantum numbers \cite{liao2}. 
Moreover, Lepton Flavor Violating (LFV) as well as Conserving (LFC) interactions are possible. 
The unparticle physics based on these interactions has rich 
phenomenological implications, and it has been studied by many authors for 
collider signature of unparticles \cite{collider}, 
neutral meson mixing system \cite{mesonmix}, 
muon anomalous magnetic moment $(g-2)$ \cite{cheung,liao,luo,chen2,hur},
LFV processes \cite{aliev,choudhury, ding,giffels,lu,iltan,iltan2}, and so on. 

Experimental results \cite{Bennett:2006fi} for muon anomalous magnetic moment $a_{\mu}=(g-2)_{\mu}/2$ are a promising hint of new physics beyond the SM. 
It is well known that between the experimental value and the SM prediction there is difference \cite{pdg,miller}
\be
\Delta a_{\mu}=a_{\mu}^{\mbox{exp}}-a_{\mu}^{\mbox{SM}}
=29.5(8.8) \times 10^{-10},
\label{g-2}
\ee
with a discrepancy of $3.4 \sigma$. 
There have been many attempts to explain this discrepancy by new physics 
(see \cite{miller,cz,jege} and references therein). 

In this paper, we investigate the muon $g-2$ and 
LFV tau decays mediated by scalar unparticles. 
If there are LFC and LFV couplings of unparticles with charged leptons, 
these couplings contribute to $g-2$ at one-loop level by 
muon loop from LFC $\mu\mu\U$ coupling \cite{cheung,liao,luo,chen2,hur} and 
tau loop from LFV $\tau\mu\U$ coupling.
These couplings generate LFV tau decay processes $\tau \to 3 \mu$ at tree level 
and $\tau \to \mu \gamma$ 
at one-loop level as well.
If the discrepancy Eq.~(\ref{g-2}) is saturated by unparticles, 
one can constrain the coupling constants and the scaling dimension $d_{\U}$
without conflicting the experimental bound on LFV tau decays. 

This paper is organized as follows: in Section~\ref{sec:unparticles}, 
we give a brief introduction of unparticles. 
In Section~\ref{sec:g-2}, unparticle mediated muon $g-2$ is studied. Section~\ref{sec:taudecay} is devoted to LFV tau decay.  
We find that there exists a consistent region of the coupling constants 
and scaling dimension $d_{\U}$ that is compatible with 
both muon $g-2$ and LFV tau decay processes. We conclude in Section~\ref{sec:conclusions}.

\section{Unparticles}
\label{sec:unparticles}

Unparticles are scale invariant objects originating from a hidden $\BZ$ sector with a
non-trivial infrared fixed point. This $\BZ$ sector is assumed to interact with the SM sector 
by exchanging very heavy particles at a high scale $M_{\U}$. The effective operator of that interaction has the form 
\be
\frac{1}{M_{\U}^{d_{\rm{SM}}+d_{\BZ}-4}}{\cal O}_{\rm{SM}}{\cal O}_{\BZ},
\ee
where ${\cal O}_{\rm{SM}(\BZ)}$ is an operator constructed by fields of the SM $(\BZ)$ sector with mass dimension $d_{\rm{SM}(\BZ)}$. Renormalization effects in the $\BZ$ sector induce dimensional transmutation at the scale 
$\Lambda_{\U}\sim 1~\TeV$. Below this scale, $\BZ$ fields match onto unparticle operators ${\cal O}_{\U}$, 
and effective interactions with the SM sector are written as 
\be
\frac{C_{\U}\Lambda_{\U}^{d_{\BZ}-d_{\U}}}{M_{\U}^{d_{\rm{SM}}+d_{\BZ}-4}}{\cal O}_{\rm{SM}}{\cal O}_{\U}
= \frac{\lambda}{\Lambda_{\U}^{d_{\rm{SM}}+d_{\U}-4}}{\cal O}_{\rm{SM}}{\cal O}_{\U}, 
\ee
where $C_{\U}$ is a coefficient fixed by the matching condition and $\lambda=C_{\U}(\Lambda_{\U}/M_{\U})^{d_{\rm{SM}}+d_{\BZ}-4}$. Although in principle the form of unparticle operator ${\cal O}_{\U}$ is determined by the theory in the hidden sector, the latter is yet unknown, and only Lorentz invariance constrains 
the unparticle operators. 

In this paper we consider LFV interactions between SM fields and unparticles 
of scalar $(S)$ and pseudo-scalar $(P)$ type. 
\be
{\cal L}=\frac{\lambda^S_{ij}}{\Lambda_{\U}^{d_{\U}-1}}\bar \ell_i \ell_j O_{\U}
+\frac{\lambda^P_{ij}}{\Lambda_{\U}^{d_{\U}-1}}\bar \ell_i i \gamma^5 \ell_j O_{\U},  
\label{int}
\ee
where $\ell_i (i=e,\mu,\tau)$ denotes charged lepton of $i$th generation, and we assume 
unparticle scale $\Lambda_{\U}=1~\TeV$ throughout this paper. The coupling constant $\lambda^S_{ij}$ and 
$\lambda^P_{ij}$ are assumed to be real. 

Propagators of scalar unparticle of momentum $P$ is derived from the 
principle of scale invariance as \cite{georgi,cheung}  
\be
\frac{i A_{d_{\U}}}{2 \sin d_{\U}\pi}\left( -P^2-i \epsilon\right)^{d_{\U}-2},
\ee
where the normalization factor $A_{d_{\U}}$ is 
\be
A_{d_{\U}}=\frac{16 \pi^{5/2}}{(2 \pi)^{2 d_{\U}}}
\frac{\Gamma(d_{\U}+1/2)}{\Gamma(d_{\U}-1)\Gamma(2 d_{\U})}.
\ee
In this paper, we consider only the region $1<d_{\U}<2$ which comes from 
unitarity condition ($1<d_{\U}$) \cite{georgi, mack, comments} and convergence condition ($d_{\U}<2$).

\section{Muon Anomalous Magnetic Moment}
\label{sec:g-2}

\begin{figure}[t]
\unitlength=1mm
\begin{picture}(70,50)
\includegraphics[width=7cm]{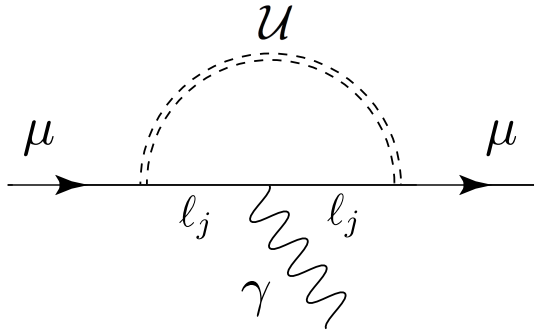}
\end{picture}
\caption{One-loop diagram of muon $g-2$ mediated by unparticles $\U$.}
\label{g2}
\end{figure}

In this section we consider unparticle contributions to muon $g-2$ by both 
LFC and LFV interactions given in \eqref{int}. 
If we assume that unparticles explain the difference $\Delta a_{\mu}$ of between experimentally measured muon $g-2$ and its SM prediction \eqref{g-2}, this condition restricts 
the possible range of the couplings. 
New contributions to $\Delta a_{\mu}$ by (pseudo)scalar unparticles are generated 
by one-loop diagram (\figref{g2}), and 
the results are 
\be
\Delta a_{\mu}^S&=&-\sum_{j=e,\mu,\tau}\frac{\left|\lambda^S_{\mu j}\right|^2}{8 \pi^2}\left( \frac{m_{j}^2}{\Lambda_{\U}^2}\right)^{d_{\U}-1}{\cal Z}_{d_{\U}}\sqrt{r_j}
\int_0^1 dz F^S(z,d_{\U},r_j), 
\label{s}\\
\Delta a_{\mu}^P&=&+\sum_{j=e,\mu,\tau}\frac{\left|\lambda^P_{\mu j}\right|^2}{8 \pi^2}\left( \frac{m_{j}^2}{\Lambda_{\U}^2}\right)^{d_{\U}-1}{\cal Z}_{d_{\U}}\sqrt{r_j}
\int_0^1 dz F^P(z,d_{\U},r_j),
\label{p}
\ee
where $j=(e,\mu, \tau)$ denotes the flavor of internal charged leptons, 
$r_j=m_{\mu}^2/m_j^2,~{\cal Z}_{d_{\U}}=A_{d_{\U}}/(2 \sin d_{\U}\pi)$ and 
functions under Feynman parameter integrals are defined as
\be
F^S(z,d_{\U},r_j)&=&z^{1-d_{\U}}(1-z)^{d_{\U}}(1+\sqrt{r_j}z)(1-r_j z)^{d_{\U}-2},
\label{fs}\\
F^P(z,d_{\U},r_j)&=&z^{1-d_{\U}}(1-z)^{d_{\U}}(1-\sqrt{r_j}z)(1-r_j z)^{d_{\U}-2}.
\label{fp}
\ee
Contribution from pseudoscalar interactions $\Delta a_{\mu}^P$ is obtained by 
replacing $m_j$ with $-m_j$ in $\Delta a_{\mu}^S$ from the chirality structure. 
One can easily verify that Eqs.~(\ref{s})--(\ref{p}) reduce to the formulae of \cite{liao} 
in the case of flavor-blind interactions when $r_j=1$. 
 
The contribution from scalar interactions $\Delta a_{\mu}^S$ to \eqref{g-2} is positive for all $j = e, \mu, \tau$. The contribution from pseudoscalar interactions $\Delta a_{\mu}^P$ has the same sign as $\Delta a_{\mu}^S$ for $j = e$ because of $r_{j=e}\gg 1$. However, for $j = \mu, \tau$ the contribution of $\Delta a_{\mu}^P$ to muon $g-2$ is negative. Therefore, we assume that pseudo-scalar couplings $\lambda_{\mu j}^{P}=0,~(j=\mu, \tau)$ in the following analysis. 
 The treatment of this $(\mu,e)$ LFV coupling is discussed below.     
 
\begin{figure}[t]
\unitlength=1mm
\begin{picture}(70,50)
\includegraphics[width=7cm]{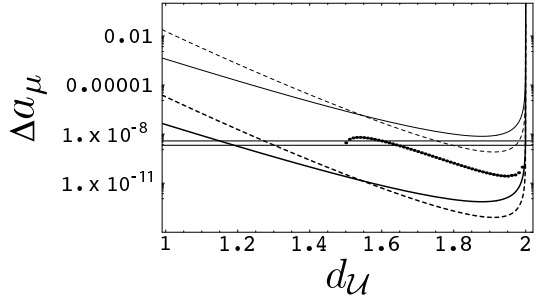}
\end{picture}
\caption{$\Delta a_{\mu}$ from the couplings with scalar unparticles as a function of the scaling 
dimension $d_{\U}$ with $\Lambda_{\U}=1~\TeV$. 
The solid, thick-solid, dashed, thick-dashed and dotted curves correspond to 
$(\lambda^S_{\mu \tau},\lambda^S_{\mu \mu},\lambda^S_{\mu e} )=(1,0,0),(10^{-4},0,0),(0,1,0),
(0,10^{-4},0)$ and $(0,0,1)$, respectively. 
All curves diverge at $d_{\U}=2$. 
The horizontal lines are the upper and lower value of \eqref{g-2}.}
\label{amm}
\end{figure}

 \figref{amm} shows $\Delta a_{\mu}$ as a function of $d_{\U}$ calculated from \eqref{s} with 
various values of $\lambda^S_{\mu \tau,\mu\mu,\mu e}$ with $\Lambda_{\U}=1~\TeV$. 
The solid, thick-solid, dashed, thick-dashed and dotted curve correspond to 
$(\lambda^S_{\mu \tau},\lambda^S_{\mu \mu},\lambda^S_{\mu e} )=(1,0,0),(10^{-4},0,0),(0,1,0),
(0,10^{-4},0)$ and $(0,0,1)$, respectively. The horizontal lines represent upper and 
lower value of \eqref{g-2}. 
From finite values of $\Delta a_{\mu}$ at $d_{\U}=1$, they are decreasing with 
larger $d_{\U}$, but diverge at $d_{\U}=2$, while the dotted curve for $d_{\U}<1.5$ is not plotted 
because it becomes negative in that region. 
Contribution of $\lambda^S_{\mu e}$ (dotted curve) is negative for $d_{\U}<1.5$ and below the 
experimental value for almost all region of $d_{\U}>1.5$ even if $\lambda^S_{\mu e}=1$. 
Moreover, this $(\mu,e)$ LFV coupling must be suppressed by experiments of 
$\mu \to e\gamma$ and $\mu-e$ conversion in nuclei \cite{ding}, and 
$\mu \to 3e$ decay process \cite{aliev} ($\lambda^S_{\mu e}=0$ is consistent with these 
experiments). In fact, at $d_{\U}=1.55$ which gives the largest contribution to $\Delta a_{\mu}$, 
upper bound of $\lambda_{\mu e}^S$ is  
$10^{-2}$ from $\mu \to e \gamma$ and $10^{-4}$ from $\mu \to 3e$ for 
$\lambda_{\mu\mu}=\lambda_{ee}=10^{-4}$.  
Such small (or vanishing) $(\mu,e)$ LFV coupling can not play 
significant role here. Therefore, we neglect this coupling in the 
following analysis.   

\begin{figure}[t]
\unitlength=1mm
\begin{picture}(70,50)
\includegraphics[width=7cm]{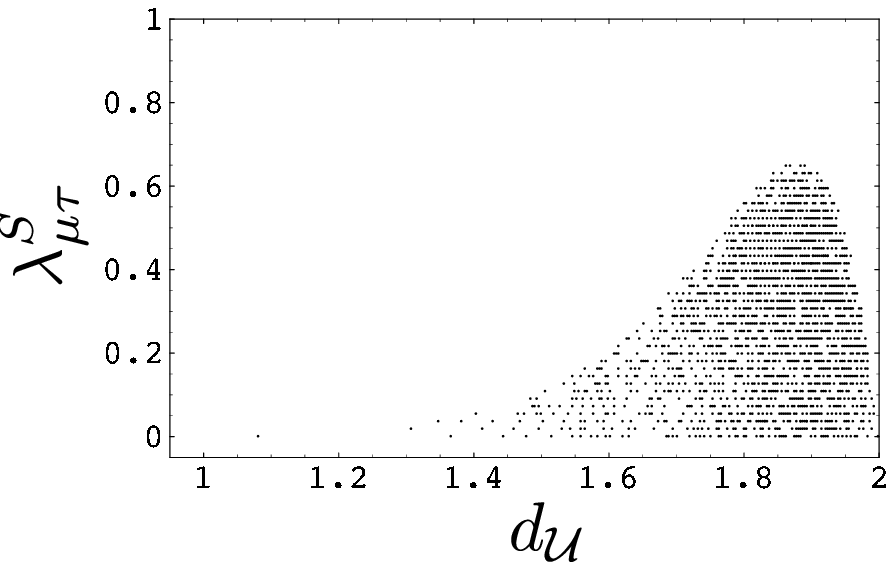}
\end{picture}
\hspace{1cm}
\begin{picture}(70,50)
\includegraphics[width=7cm]{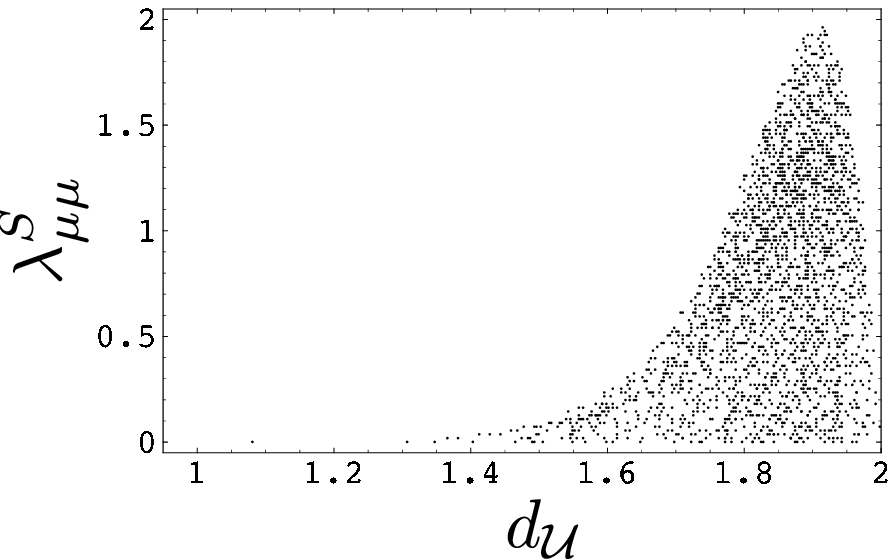}
\end{picture}
\caption{Allowed region of $\lambda^S_{\mu \tau}$ (left) and $\lambda^S_{\mu \mu}$ (right) 
from the condition of $\Delta a_{\mu}$.}
\label{mtmm}
\end{figure}

\begin{figure}[t]
\unitlength=1mm
\begin{picture}(70,50)
\includegraphics[width=7cm]{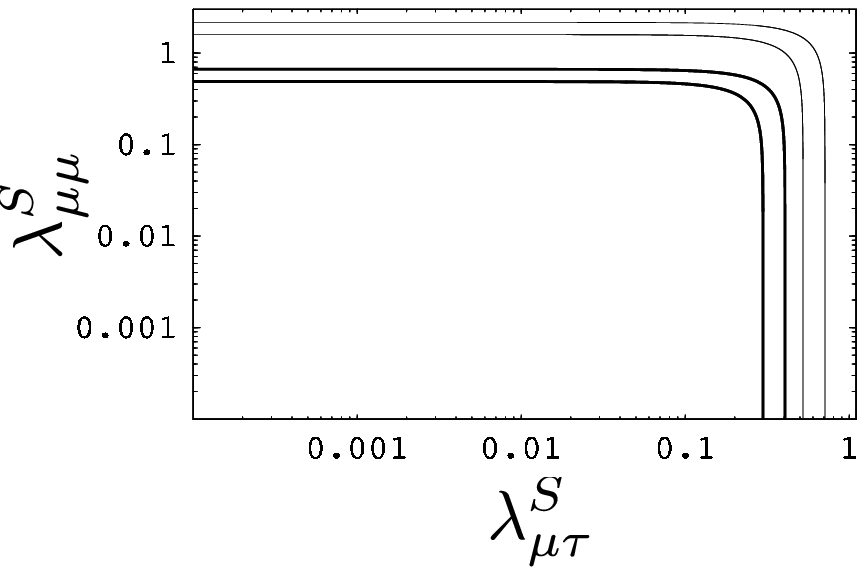}
\end{picture}
\hspace{1cm}
\begin{picture}(70,50)
\includegraphics[width=7cm]{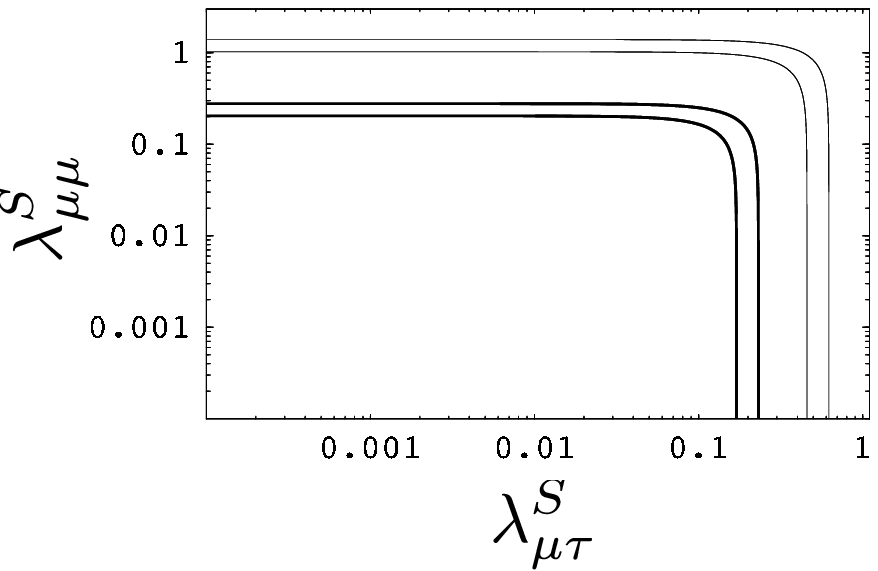}
\end{picture}
\caption{Allowed region of the coupling constants in the 
$\lambda^S_{\mu\tau}-\lambda^S_{\mu \mu}$ plane with $d_{\U}=(1.7,1.9)$ (left) and 
$d_{\U}=(1.6,1.8)$ (right). 
In the left panel, the region surrounded by thick-solid (solid) curves 
corresponds to $d_{\U}=1.7(1.9)$, and similarly for the right panel with $d_{\U}=1.6(1.8)$. }
\label{lamlamg2}
\end{figure}

\figref{mtmm} shows the consistent region of $\lambda^{S}_{\mu \tau}$ (left) and 
$\lambda^{S}_{\mu \mu}$ (right) at $\Lambda_{\U}=1~\TeV$ under the assumption that  
both LFC and LFV couplings simultaneously contribute to $\Delta a_{\mu}$, and it is 
in the bound of \eqref{g-2}. For $\lambda_{\mu \tau}^S$, 
$\lambda_{\mu\mu}^S$ is only a free parameter, and {\em vice versa}.   
For both couplings, $\lambda^S$s have to 
be small for the region of small $d_{\U}$, but they can be of ${\cal O}(1)$ for relatively large $d_{\U}$. 
They must be extremely small when $d_{\U}$ is closer to 2, 
and there is no solution at $d_{\U}=2$ because $\Delta a_{\mu}$ diverges at this point. 
Since we have assumed that both $\lambda^{S}_{\mu \tau}$ and 
$\lambda^{S}_{\mu \mu}$ 
contribute, these allowed regions are not independent of each other. 
The relation between these two couplings are shown in \figref{lamlamg2}. 
These figures represent the allowed region of $\lambda^S$s in the 
$\lambda^S_{\mu \tau}-\lambda^S_{\mu\mu}$ plane with $d_{\U}=(1.7,1.9)$ (left) 
and $d_{\U}=(1.6,1.8)$ (right). In the left panel, the region surrounded by thick-solid (solid) 
curves correspond to $d_{\U}=1.7(1.9)$, and similarly for the 
right panel. The ``slice'' of this area goes inside with decreasing $d_{\U}$, and 
we emphasize that both or at least one coupling 
of $\lambda^S_{\mu\tau(\mu\mu)}$ have to be of ${\cal O}(0.1 \ldots 1)$ 
for large $d_{\U} (\gsim 1.6)$. These relatively large couplings may raise a problem 
on LFV tau decay that we study in the next section.   



\section{LFV Tau decay}
\label{sec:taudecay}

In this section, we investigate the LFV $\tau$ decay processes generated by 
the same unparticle interactions as those of $\Delta a_{\mu}$. 
We discuss $\tau \to 3 \mu$ and $\tau \to \mu \gamma$. 
Since the constraints of couplings $\lambda^S_{\mu \tau(\mu \mu)}$ obtained from the consistency of 
 $\Delta a_{\mu}$ in the previous section tolerate large LFV couplings, 
 our next task is to make certain that this does not conflict with the experimental bound of 
 LFV tau decays  \cite{tauexp,pdg}
 \be
BR(\tau \to 3 \mu)<3.2 \times 10^{-8},~~BR(\tau \to \mu \gamma)<6.8 \times 10^{-8}. 
\label{bound}
 \ee  
Here, we will find regions of couplings which are consistent with both $\Delta a_{\mu}$ and LFV tau decays. 

\begin{figure}[t]
\unitlength=1mm
\begin{picture}(70,50)
\includegraphics[width=7cm]{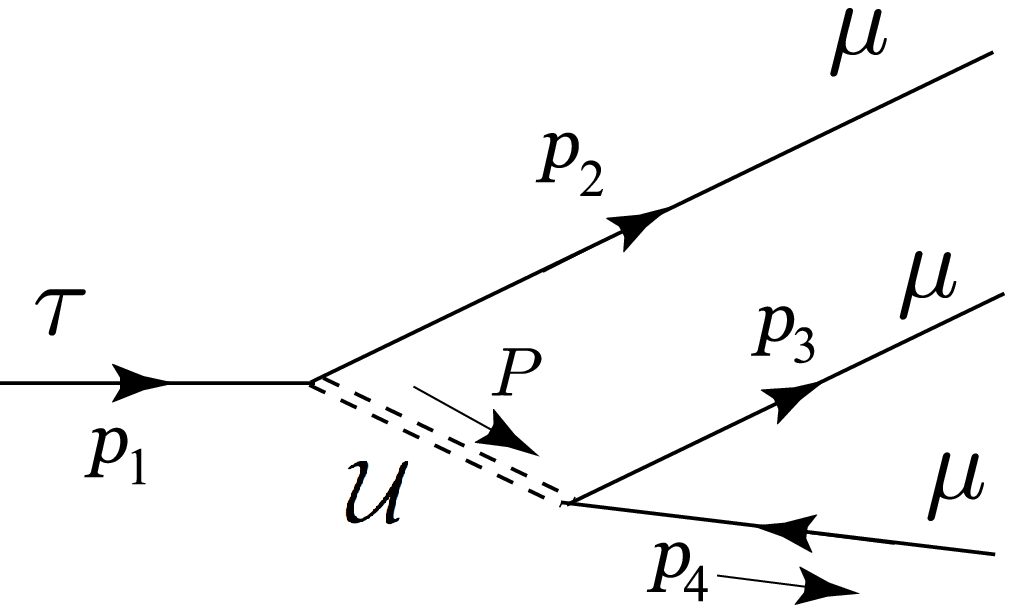}
\end{picture}
\hspace{2cm}
\begin{picture}(70,50)
\includegraphics[width=7cm]{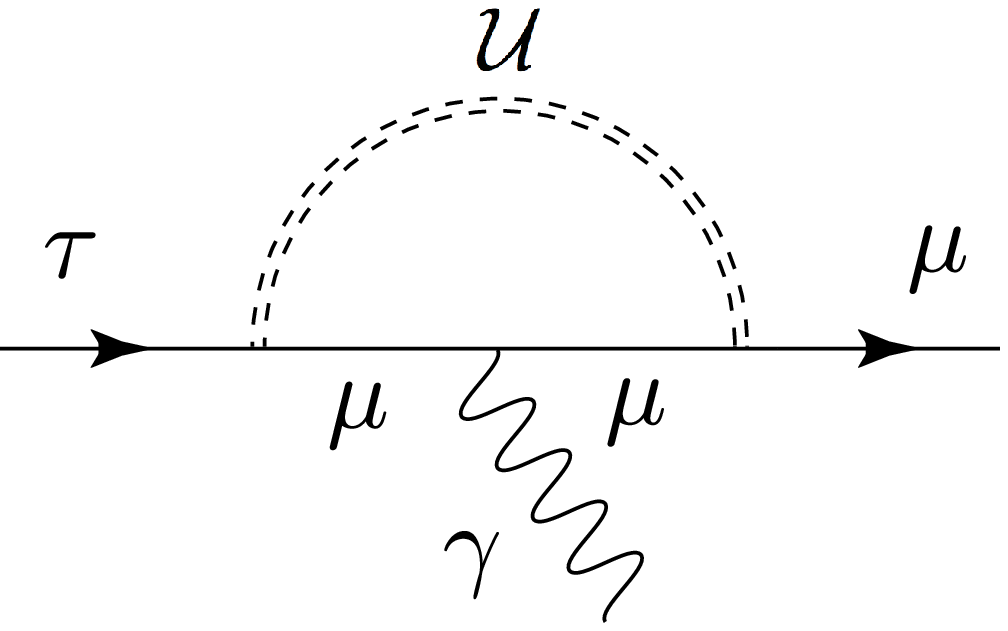}
\end{picture}
\caption{Tree-level diagram of LFV $\tau \to 3 \mu$ decay mediated by unparticles (left) and 
$\tau \to \mu \gamma$ at one-loop level (right). 
For  $\tau \to 3 \mu$, there also exists the $u$-channel diagram by exchanging external muons of momenta $p_2$ and $p_3$.}
\label{tmmm}
\end{figure}

First we consider $\tau \to 3 \mu$ LFV tau decay \cite{aliev, giffels}. This decay mode mediated by 
unparticle operators occurs at tree-level (\figref{tmmm}, left). 

The decay rate of $\tau \to 3 \mu$ derived in \cite{aliev} is
\be
\frac{d \Gamma}{ds \sin \theta d\theta}&=&\frac{1}{2^9 \pi^3}\frac{1}{\sqrt{s}}
\sqrt{\left( 1-\frac{(m_{\tau}-m_{\mu})^2}{s}\right)\left( 1-\frac{(m_{\tau}+m_{\mu})^2}{s}\right)}
\sqrt{1-\frac{4 m_{\mu}^2}{s}}\sum_{\mathrm{spin}}\left|{\cal M}\right|^2,
\ee
where $\theta$ is angle between three-momenta {\bf p}$_1$ and {\bf p}$_4$, $s=(p_1-p_2)^2$ and its integral range is $4 m_{\mu}^2 \le s \le (m_{\tau}-m_{\mu})^2$. 
Amplitude ${\cal M}$ is  
\be
\sum_{\mathrm{spin}}\left|{\cal M}\right|^2&=&
4 \left|\lambda_{ \mu \tau}^S \right|^2 \left|\lambda_{\mu \mu}^S \right|^2
\left[ 4(p_1 \cdot p_2)(p_3 \cdot p_4)|F_1|^2+ 4(p_1 \cdot p_3)(p_2 \cdot p_4)|F_2|^2 \right.\nn \\
&-& \left.2 \left\{ (p_1 \cdot p_2)(p_3 \cdot p_4)+(p_1 \cdot p_3)(p_2 \cdot p_4)
-(p_1 \cdot p_4)(p_2 \cdot p_3)\right\}\mbox{Re}(F_1F_2^*)\right], 
\ee
where 
\be
F_1&=&\frac{{\cal Z}_{d_{\U}}}{\Lambda_{\U}^{2(d_{\U}-1)}}\left( -(p_1-p_2)^2-i\epsilon\right)^{d_{\U}-2},\nn \\
F_2&=&\frac{{\cal Z}_{d_{\U}}}{\Lambda_{\U}^{2(d_{\U}-1)}}\left( -(p_1-p_3)^2-i\epsilon\right)^{d_{\U}-2}.
\ee

LFV unparticle interactions can generate other $\tau \to 3 \ell$ decay processes in general, and $\tau \to e \mu \mu$ is the one of them which contains the coupling  responsible for muon $g-2$. This process contains 
$\lambda^S_{\mu e}$ as well as 
$\lambda^S_{\mu \tau}$. However, as mentioned in the previous section, we have neglected 
this $(\mu,e)$ LFV coupling because it is strongly suppressed by other LFV processes. 
   
$\tau \to \mu \gamma$ process is the other decay mode by the same unparticle interactions 
at one-loop level (\figref{tmmm}, right). 
Decay Rate of this process is \cite{ding}
\be
\Gamma=\frac{m_{\tau}^3}{8 \pi}|{\cal A}|^2
\ee 
where the amplitude ${\cal A}$ is given by
\be
{\cal A}&=&-\sum_{j=e,\mu, \tau}\frac{i e}{(4 \pi)^2}\lambda^S_{\mu j}\lambda^S_{j \tau}{\cal Z}_{d_{\U}}
\frac{1}{\left(\Lambda_{\U}^2\right)^{d_{\U}-1}}G^S_j (d_{\U}),
\label{tmgamp}
\ee
and functions $G^S_j(d_{\U})$ are
\be
G^S_j (d_{\U})&=&\int dx dy dz \delta(x+y+z-1)z^{1-d_{\U}} \nn \\
&~&\left[ -x z m_{\tau}^2-y z m_{\mu}^2+(x+y)m_j^2\right]^{d_{\U}-2} 
\left[ x z m_{\tau}+y z m_{\mu}+(x+y)m_j\right]. 
\ee
While three leptons $j=(e,\mu,\tau)$ can 
exist in the loop, we consider only the case of muon virtual particle because we are interested in 
tau decays generated by the same couplings as those of muon $g-2$. 
Contributions from other virtual particles depend on different combination of 
$\lambda^S$, such as $\lambda^S_{\tau e} \lambda^S_{\mu e}$ for electron loop and 
 $\lambda^S_{\tau \tau} \lambda^S_{\mu \tau}$ for tauon loop. 
Moreover, $\tau \to e\gamma$ is also possible LFV tau decay mediated by unparticles. 
However this process also contains unknown, or more suppressed parameters 
$\lambda^S_{ee,\tau e, \tau\tau}$. These couplings 
 may be constrained by other processes, but we neglect these here because all of them 
must be small or can be zero.  
 
If there are couplings of unparticle with photons \cite{iltan2}, 
\be
\frac{1}{\Lambda_{\U}^{d_{\U}}}\left( \lambda_{\gamma}F_{\mu \nu}F^{\mu \nu}+
\lambda_{\tilde \gamma}\tilde F_{\mu \nu}F^{\mu \nu}\right)O_{\U}
\ee 
these operators also generate $\tau \to \mu \gamma$ at one-loop level. 
However, these operators are more suppressed by the factor $m_{\tau}/\Lambda_{\U}\sim 10^{-3}$ than \eqref{tmgamp}, and contain unknown parameters 
$\lambda_{\gamma (\tilde \gamma)}$. Therefore, we again neglect these interactions. 

In the next subsection, we perform numerical calculation of these LFV tau decay processes, and verify that there exist regions which do not conflict with experiments. 

\subsection*{Numerical Calculation}

\begin{figure}[t]
\unitlength=1mm
\begin{picture}(70,50)
\includegraphics[width=7cm]{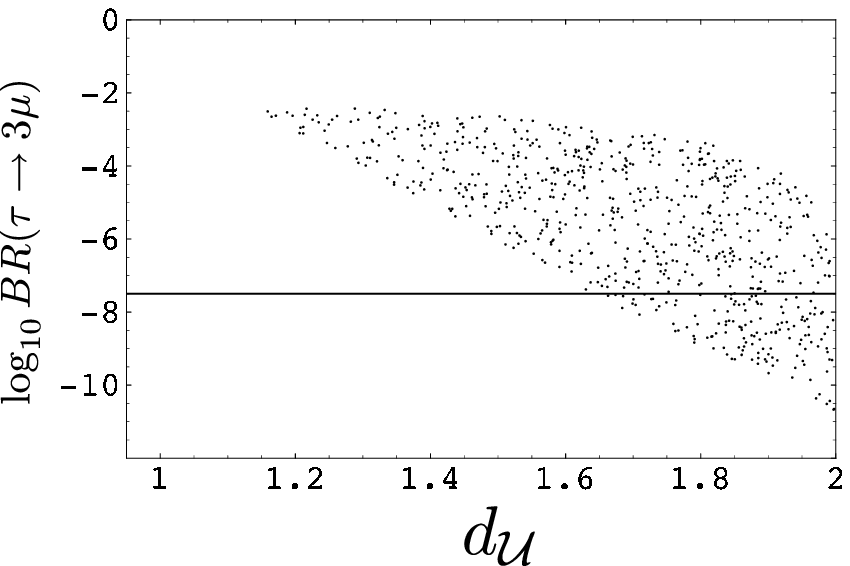}
\end{picture}
\hspace{1cm}
\begin{picture}(70,50)
\includegraphics[width=7cm]{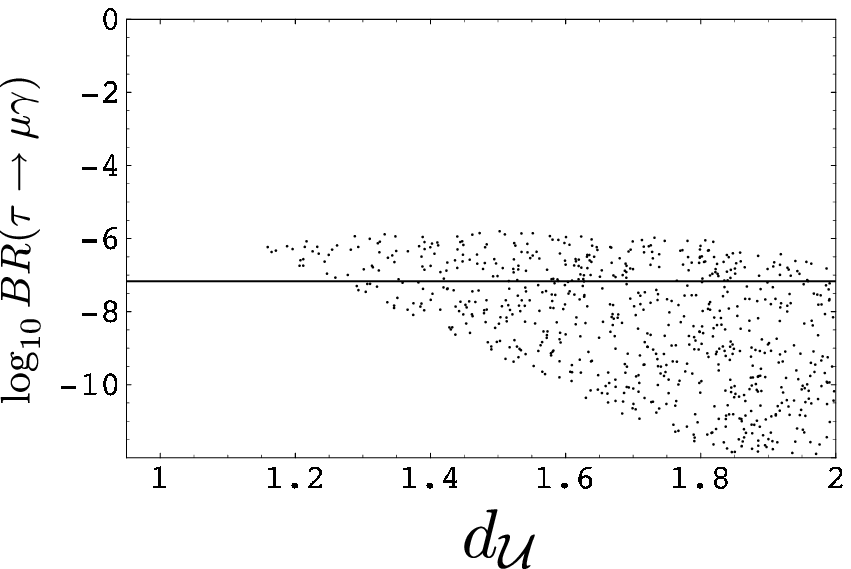}
\end{picture}
\caption{Branching ratio of $\tau \to 3 \mu$ (left) and $\tau \to \mu \gamma$ (right) 
by scalar unparticle operators with $\lambda^S_{\mu \mu,\mu\tau }> 0.001$. 
The horizontal line is experimental upper bound of \eqref{bound}.  
There exist solutions below the experimental bound for 
$d_{\U} \raise0.3ex\hbox{$\;>$\kern-0.75em\raise-1.1ex\hbox{$\sim\;$}} 1.6$.}
\label{td}
\end{figure}

Now we are ready to find whether unparticle can explain muon $g-2$ without conflicting with LFV tau decay processes. 
\figref{td} show the BR of $\tau \to 3 \mu$ (left) and $\tau \to \mu \gamma$ (right) 
as a function of $d_{\U}$. In these figures, 
$\lambda^S_{\mu \tau}$ and $\lambda^S_{\mu \mu}$ are generated 
independently and randomly in the region allowed by $g-2$ experiment shown 
in \figref{amm} assuming $\lambda^S_{\mu \tau, \mu \mu}>0.001$. 
The horizontal lines represent experimental bound \eqref{bound}. 
From the figures, one can see that some sets of 
$(\lambda^S_{\mu\tau},\lambda^S_{\mu\mu})$ give the BR below the experimental bound 
of $\tau \to 3 \mu$ if $d_{\U}\gsim1.6$, 
while it is enough suppressed for almost all $d_{\U}$ for 
the one-loop process $\tau \to \mu \gamma$. 
These results are not changed for larger 
$\Lambda_{\U}$, because the dependence of $\lambda^S/\Lambda_{\U}^{d_{\U}-1}$ is 
the same for all phenomena.

\begin{figure}[t]
\unitlength=1mm
\begin{picture}(70,50)
\includegraphics[width=7cm]{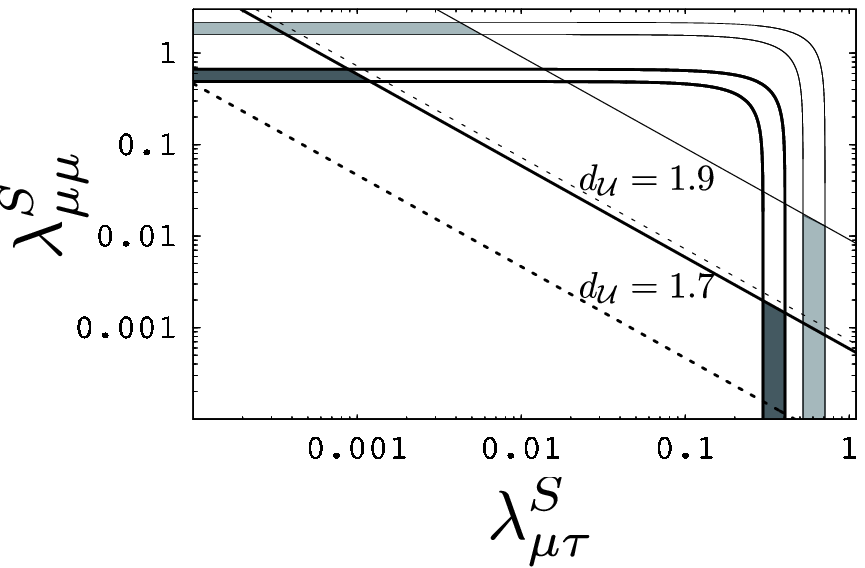}
\end{picture}
\hspace{1cm}
\begin{picture}(70,50)
\includegraphics[width=7cm]{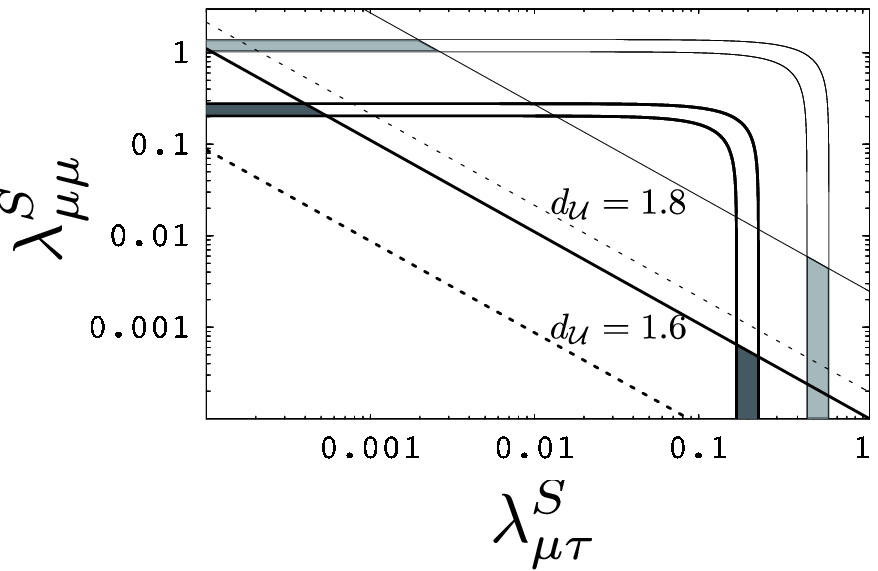}
\end{picture}
\caption{Consistent region in the $\lambda^S_{\mu \tau}-\lambda^S_{\mu \mu}$ plane 
at $d_{\U}=(1.7,1.9)$ (left) and $d_{\U}=(1.6,1.8)$ (right). Left lower areas of each diagonal line representing  
the allowed region from $\tau \to 3 \mu$ experiment are superimposed on \figref{lamlamg2}, and 
dotted lines are obtained from expected future experiments. Shaded areas 
represent the combined allowed region of all present experiments.  }
\label{lam2}
\end{figure}

\figref{lam2} shows the final region in the $\lambda^S_{\mu \tau}-\lambda^S_{\mu \mu}$ plane at $d_{\U}=(1.7,1.9)$ (left) and $d_{\U}=(1.6, 1.8)$ (right). 
In the left figure, thick curves and lines correspond to $d_{\U}=1.7$ and 
thin ones to $d_{\U}=1.9$, and similar for the right figure. 
Left lower areas of each diagonal solid line representing  
the allowed region from $\tau \to 3 \mu$ experiment \eqref{bound} are superimposed on \figref{lamlamg2}. 
Dark and light shaded areas 
represent the combined allowed region of all experiments for $d_{\U}=1.7(\mbox{left}),1.6(\mbox{right})$ and 
$d_{\U}=1.9(\mbox{left}),1.8(\mbox{right})$, respectively.
Future experiments such as super B factory \cite{kekb} will have sensitivity 
up to $2 \times 10^{-10}$ for $\tau \to 3 \mu$ and $2 \times 10^{-9}$ for $\tau \to \mu \gamma$, 
and dotted lines are obtained from $BR(\tau\to 3 \mu)<2 \times 10^{-10}$.  
 
Comparing the \figref{lamlamg2}, the regions in which both couplings are large vanish, 
and those in which either of them is large remain. These regions are compatible with 
both muon $g-2$ and LFV tau decay processes. This means either of the couplings 
has to be of ${\cal O}(0.1 \ldots 1)$ for the case of large $d_{\U}$.
In the case of both the couplings are small, which is favored in the point of view of LFV tau decay, 
the scaling dimension $d_{\U}$ has to be small in order to obtain appropriate value of muon $g-2$. 
Future LFV tau decay experiments will restrict the allowed regions.     

The situation $\lambda^S_{\mu \tau}=0$ is also a solution. 
The coupling $\lambda^S_{\mu \mu}$ can give a desired value of $\Delta a_{\mu}$ even if 
$\lambda^S_{\mu \tau}=0$, and in this case LFV tau decays of our consideration can not occur. 

We conclude that LFV coupling $\lambda^S_{\mu \tau}$ does not have to be zero or extremely suppressed, 
it can be of ${\cal O}(0.1\ldots 1)$, if  $\lambda^S_{\mu \mu}\lsim 10^{-2}$  and $d_{\U}\gsim 1.6$.


\section{Conclusions}
\label{sec:conclusions}

We have studied muon anomalous magnetic moment and lepton flavor violating tau decay $\tau \to 3 \mu$ and $\tau \to \mu \gamma$ generated by scalar unparticle interactions. Since the principle to determine unparticle 
interactions is still unknown, both lepton flavor violating and conserving interactions may exist, and  
these interactions can 
simultaneously generate both phenomena. 
We have found that scalar unparticles explain the discrepancy of experimental value of 
muon $g-2$ from the Standard Model prediction, 
without conflicting with the experimental bound of LFV tau decay processes. 
When either LFV or LFC coupling vanishes, muon $g-2$ is easily generated and 
LFV tau decay can not occur in any value of the scaling dimension $d_{\U}$. 
On the other hand, when both couplings exist, these couplings and the scaling dimension are constrained 
by LFV tau decay. 
In the case of large scaling dimension $(d_{\U}\gsim 1.6)$, 
LFV coupling $\lambda^S_{\mu \tau}$ need not be small.  
It can be of ${\cal O}(1)$ if 
LFC coupling $\lambda^S_{\mu \mu}$ is enough small, 
and \textit{vice versa}. 

\vspace{0.5cm}
\noindent
{\large \bf Acknowledgments }\\
The authors would like to thank M. Raidal for useful discussions. 
This work is supported by the ESF grant No. 6190, and postdoc
contract 01-JD/06 (Y.K.).                                                                           
\bibliographystyle{unsrt}

\end{document}